# The microstrain-doping phase diagram of the iron pnictides: heterostructures at atomic limit

Alessandro Ricci,  Nicola Poccia, Gabriele Ciasca, Michela Fratini and Antonio Bianconi

*Department of Physics, Sapienza University of Rome, P. Aldo Moro 2, 00185 Roma, Italy*

## Abstract

The 3D phase diagram of iron pnictides where the critical temperature depends on charge density and microstrain in the active FeAs layers is proposed. The iron pnictides superconductors are shown to be a practical realization of a heterostructure at the atomic limit made of a superlattice of FeAs layers intercalated by spacers layers. We have focussed our interest on the $A_{1-x}B_xFe_2As_2$ (122) families and we show that FeAs layers have a tensile microstrain due to the misfit strain between the active layers and the spacers. We have identified the critical range of doping and microstrain where the critical temperature gets amplified to its maximum value.

## 1. Introduction

Understanding the quantum mechanism that allows a macroscopic quantum condensate, a superfluid or a superconductor, to resist the decoherence effects of high temperature is a major topic in condensed matter, quantum computing and in the search for quantum mechanisms in the living cell. The evidence that it is possible to achieve a quantum condensate of fermions at high temperatures is provided by the so called high $T_c$ superconductors. Recently the discovery of the new FeAs high $T_c$ superconducting multilayers [1] provides a test for the reliability of the models already proposed to explain high $T_c$ superconductivity (HTcS) in others complex materials.

The structure of the iron pnictides superconductors is made of a superlattice of $\left[FeAs\right]_{\infty}^{-Q+\delta}$ with Q=1, layers intercalated by spacers (oxide layers like $\left[LnF_yO_{1-y}\right]_{\infty}^{+Q-\delta}$ or $\left[LnO_{1-y}\right]_{\infty}^{+Q-\delta}$ in the "1111" family or metallic atomic layers $\left[(A_{1-x}^{+2}B_x^{+1})_{1/2}\right]_{\infty}^{+Q-\delta}$ in the "122" family) therefore they represent practical realizations of a heterostructure at the atomic limit (HsAL) that was described to be the essential material architecture for the emergence of HTcS [2].

In fact, different realizations of HsAL are the cuprate superconductors, where the $\left[CuO_2\right]_{\infty}^{-2+\delta}$ active layers are intercalated by spacers like the $\left[La_2O_{2+y}\right]_{\infty}^{+2-\delta}$ block layers [3,4], and magnesium diborides where the $\left[B_2\right]_{\infty}^{-3+\delta}$ active layers are intercalated by spacers like $\left[Al_{1-x}Mg_x\right]_{\infty}^{+3-\delta}$ layers [5,6]. Experimental results on pnictides have shown that the chemical potential should be driven into a particular point of the electronic phase diagram by controlling the charge density [1], the pressure [7], and the spacer material in order to reach the superconducting phase with $T_c$ as high as 55 K [8]. Therefore it now well established that all known HTcS superconductors have the HsAL material architecture. In these superlattices the metallic active layers have strong covalent bonds (the $CuO_2$ layers, the graphene-like $B_2$ monolayers, or the molecular $FeAs_{4/4}$ layers) intercalated by spacers made of different materials with a different electronic structure (fcc rocksalt oxide layers like

La$_{2-x}$Sr$_x$O$_{2+y}$ [3, 4], or hcp metallic Mg/Al layers [5,6], and rare-earth oxide layers or atomic metallic layers pnictides "1111" or "122" family respectively).

## 2     Manganites HsAL

The same kind of architecture is also observed in manganite perovskites AMnO$_3$ that show the colossal manetoresistence (CMR). The perovskites are composite materials made of transition metal oxides and rare earth oxides AO. This structure is formed by the matching of the equilibrium (A–O) distance in the rare earth metal oxide and the equilibrium (M–O) bond length in the transition metal (M=Mn) oxide. Ideal matching occurs where $t = (A-O)/((M-O)\sqrt{2})$ the geometric tolerance factor is unity. Also these materials can be considered HSAL, made of a superlattice of $\left[ MnO_2 \right]_\infty^{-2}$ bcc layers intercalated by spacers of $\left[ LnO \right]_\infty^{-2}$ rare earth oxide fcc layers.

The CMR phenomenon occurs at particular values of charge transfer δ (called also doping) between the active layers $\left[ MnO_2 \right]_\infty^{-2+\delta}$ and the spacer layers $\left[ A_{1-x}^{+3} B_x^{+2} O \right]_\infty^{+2-\delta}$ but also for a fixed carrier concentration δ they reveal a direct relationship between the Curie temperature and the average ionic radius of the rare earth in the spacer layers i.e. the geometric tolerance factor [9].

In a superlattice of two different components the lattice mismatch is measured by the misfit strain defined as $\eta = (a_1 - a_2)/\langle a \rangle$ with $\langle a \rangle = (a_1 + a_2)/2$ where a$_1$ (a$_2$) is the lattice parameter of the first (second) component when it is well separated. Assembling the two components results in a superlattice where the lattice mismatch is accommodated by a resulting compressive (tensile) microstrain in the first (second) component or viceversa [10]. It is trivial to show that the superlattice misfit strain in the perovskites between bcc monolayers and fcc monolayers is given by $\eta = 1 - t$. Therefore the CMR phenomenon occurs at particular values of doping and misfit strain.

In high T$_c$ cuprate superconductors it was proposed that the maximum T$_c$ occurs in a particular point of the electronic structure of the HsAL where the chemical potential is tuned to a particular point changing the lattice parameters and/or the charge density in the active layers to get a shape resonance (called also Feshbach resonance) of the interband pairing in a multiband superconductor [2,4,6].

## 3. Cuprates HsAL

The microstrain in the active metallic layers has been proposed to be the conformational parameter or the physical variable for a HsAL that describes the tuning of the chemical potential to the Feshbach resonance by the lattice variations [11,12] i.e., by changing the chemical pressure or internal pressure. For doped $La_{2-x}Sr_xCuO_{2+y}$ (La214) cuprate superconductors, where the $[CuO_2]_\infty^{-2+\delta}$ active layers are intercalated by spacers $[La_{2-x}Sr_xO_{2+y}]_\infty^{+2-\delta}$, it is easy to calculate the tolerance factor, and to deduce the misfit strain, showing that the copper oxide active layers are under a compressive chemical pressure. However in cuprates that was not possible to calculate the tolerance factor and the misfit strain in other families because of the intricate structure of block layers with a plurality of cations having largely different coordination numbers.

The information on the elastic field acting on the active layers in the strained layers of a superlattice can be obtained by measuring the microstrain [11-13]. This is given by the difference of the measured lattice parameter of the first component in the assembled compound and the lattice parameter of the first component when it is isolated. This problem was in fact solved for the cuprates by measuring the compressive microstrain of the $CuO_2$ monolayer $\varepsilon = -(r - R_0)/R_0$ where $r$ is the measured average Cu-O distance in the superlattice and $R_0 = 0.197$ $nm$ is the Cu-O equilibrium distance in the well separated $[CuO_2]_\infty^{-2}$ layers. In the strained superlattice the tensile microstrain in the spacer layers is expected to have the same absolute value as the compressive microstrain in the active layers. In the case of similar elastic constants in the two components the misfit strain (measuring the internal chemical pressure) is two times the microstrain.

Therefore it was possible to put in the same phase diagram all families of cuprates where the elastic field due to the lattice mismatch is measured by microstrain and the charge density in the active layers by doping [11-13].

In these 3D phase diagrams there are regions of charge ordering, regions of phase separation, critical points and finally regions of HTcS. It was proposed that the highest $T_c$ occurs at a critical point in the 3D phase diagram where the phase separation vanishes. Recently a model has been pointed out that shows that if there is a multiband metal in the active layers in a range of values of doping and microstrain the system shows phase separation and there are critical points where the phase separation vanishes [14,15]. A similar phase diagram has been proposed for diborides [6,16].

The purpose of this work is to show that also for the new iron pnictides both the microstrain and the charge density in the $\left[FeAs\right]_{\infty}^{-Q+\delta}$ layers are essential parameters for driving the chemical potential to a particular point with optimum doping and microstrain where the Feshbach resonance is expected to amplify the critical temperature according with ref. [2,4,6].  We have focused our interest on the "122" family where the spacer layers are simple metallic atomic layers. In the undoped superlattices the active layers $\left[FeAs\right]_{\infty}^{-1}$ and $\left[FeAs\right]_{\infty}^{-0.5}$ increasing the ionic radius in the spacer suffer a tensile microstrain.  The FeAs equilibrium distance in the $\left[FeAs\right]_{\infty}^{-1}$ is 8.2 pm longer than in $\left[FeAs\right]_{\infty}^{-0.5}$ superlattices therefore the lattice of the FeAs layers show a large lattice relaxation with the variation of the formal chemical charge. From these data we can deduce the microstrain as a function of doping and we have identified the critical values of microstrain and doping where the maximum $T_c$ occurs in the "122" iron picnitides families".

# 4. Microstrain- doping phase diagram for pnicitides HsAL

In Fig. 1 the Fe-Fe distance is plotted as a function of the average ionic radius in the spacer layers for the "122" family of FeAs based pcnitides [18-21]. The systems where intercalated ions in spacer layer have the same charge are connected by a polynomial line. It is worth to notice that the two curves linking the systems with intercalated $A^+$ ions and $A^{2+}$ ions are approximately parallel [22]. We remark that there is a non linear response of the lattice as a function of the ionic radius, if it is too small (like that of lithium), i.e. below a critical radius of the ions in the spacers, it does not introduce a microstrain in the FeAs layer, therefore we take the value of the Fe-Fe distance in this regime as the unstrained distances are $R_0^{-1} = 276.35$ $pm$ and $R_0^{-0.5} = 268.05$ $pm$ in the $[FeAs]_\infty^{-1}$ and $[FeAs]_\infty^{-0.5}$ layers respectively. The microstrain of the $[FeAs]_\infty^{-1}$ layers can be easily measured. In fact these layers are made of edge sharing $FeAs_4$ tetrahedral units where the FeAs bond length remains constant therefore the chemical mismatch pressure, the misfit strain, induces only a rotation of the bonds pushing the As–Fe–As bond out of the ideal value of the tetrahedral angle 109.28° [23], where the ideal lattice parameter of the orthorhombic lattice is $a_o = \sqrt{2}a_T = 552.7$ pm. The microstrain is therefore given by $\varepsilon = (a_o/552.7 - 1) = (a_T/390.82 - 1)$ [24,25].

The figure shows that there is lattice contraction of 8.3 pm (with a constant ionic radius in the intercalated layers) only due to the variation of the effective formal charge in the FeAs layers of 0.5e. In Fig. 1 the curves of doped iron pnictides with $[(Ba_{1-x}^{+2}K_x^{+1})_{1/2}]_\infty^{+1-\delta}$, $[(Sr_{1-x}^{+2}Cs_x^{+1})_{1/2}]_\infty^{+1-\delta}$ and $[(Sr_{1-x}^{+2}K_x^{+1})_{1/2}]_\infty^{+1-\delta}$ intercalated atomic layers follow a linear line as expected for the Vegard law connecting the stoichiometric compounds. The case of spacer layers $[(Ba_{1-x}^{+2}K_x^{+1})_{1/2}]_\infty^{+1-\delta}$ is of particular interest since the average ionic radius remains nearly constant.

In Fig. 2 we have plotted the Fe-Fe distance as a function of the charge density in the FeAs layers. Decreasing the charge in the FeAs layers, we observe a decreasing of the Fe-Fe distance as we substitute $A^{+1}$ for $A^{+2}$ ions. A mesoscopic phase separation is observed in the range $0 < \delta < 0.1$ and the highest $T_c$ is reached in the range $0.1 < \delta < 0.2$ where a subtle tuning between the charge density and elastic strain field which are non trivially mixed. The variation of the unstrained Fe-Fe distance only due to the variation of the formal charge $\delta$ in the FeAs layers is indicated by the solid line $R_0(\delta) = 276.35 - 16.6$ $\delta$ pm

The microstrain is a response function of the material which measures the accommodated misfit strain between spacer and active layers. The microstrain in the active layers, defined as $\varepsilon = (R(\delta) - R_0(\delta))/R_0(\delta)$ where $R(\delta)$ is the measured Fe-Fe distance in the doped superlattice, is plotted in Fig. 3 as a function of the charge density (or doping). In the plane microstrain – doping we can see that there are regions of 1) striped phase i.e. a region of itinerant magnetic order with orthorombic lattice distortions, at doping range $\delta=0$, a phase separation region for $0<\delta<0.2$ and the shaded area indicates the maximum $T_c$ region centered at at the critical microstrain $\varepsilon_c \approx 0.12$, misfit strain $\eta_c = 2\varepsilon \approx 0.24$ and doping $\delta_c \approx 0.2$.

## 5. Semiconductor heterostructures versus superconducting HsAL

We have shown that cuprates, diborides and pnictides have a similar heterostructure at atomic limit. This architecture is found in all lamellar materials, exhibiting $HT_cS$, two repeated different layers: one which is superconducting, the other which is a spacer. The alternated layers show periodically compressive and tensile stress which is measured and named microstrain. A couple of compressive and tensile alternated layers is a typical feature of specific kind of semiconductor heterostructures named strain – balanced structures. These are pseudomorphic superlattice structures made of alternated tensile and compressive strained layer. Because of the stress balancing, the tensile/compressive coupled layers reach an in plane lattice parameter $a_m$. In order to achieve the overall zero-stress condition the alternated layer stack should be grown on a substrate with the same lattice parameter $a_m$ [26]. The presence of a periodically alternated compressive/tensile strain in the superconducting/spacers multilayer develops to balance the stress acting on the stack, as it happens in semiconducting strained-balanced superlattices. In an infinite superconducting crystal the presence of a residual stress would unlimitedly raise the total amount of elastic energy stored in the crystal, leading to a non-physical situation. Different definitions of chemical pressure are used in the condensed matter physics [10]. Chemical pressure due to the material epytaxial growth or the hydrostatic pressure are different kind of pressures that do not concern the model which here we discuss. An hydrostatic pressure variable beside doping has yet been used to describe the phase diagram of the FeAs superconductor [27].

## 6. Conclusion

Several models have been proposed in the past in order to universally describe the phenomenon of superconductivity with a high critical temperature. We have shown that both the microstrain and the charge density on FeAs layers are essential parameters for understanding the variation of the critical temperature in the FeAs compounds (122). We propose that the presence of a periodically alternated compressive/tensile strain in the superconducting/spacers multilayer could balance the stress acting on the stack, as it happens in semiconducting strained-balanced superlattices. The critical temperature Tc is controlled by both charge density and lattice effects and the maximum $T_c$ region is centered at at the critical microstrain $\varepsilon_c \approx 0.12$, misfit strain $\eta_c = 2\varepsilon \approx 0.24$ and doping $\delta_c \approx 0.2$. Furthermore we have shown a large variation of the Fe-Fe distance due only to changes of the formal chemical charge at fixed elastic strain field. In fact it is shown that the Fe-Fe distance is 8.3 pm longer for FeAs layers with formal charge Q= -1 than in layers with formal charge Q=-0.5.

**Aknowlegements:** This work was supported by European project 517039 "Controlling Mesoscopic phase separation" (COMEPHS) (2005).

# Figure Captions:

**Figure 1:** The Fe-Fe distance is plotted as a function of the average ionic radius for 122 family FeAs compounds [24, 25]. Ions with same charge are arranged along the same curve. It is worth to notice that the curves referred as $A^+$ ions and $A^{2+}$ ions are approximately parallel. Infact, for systems with charge +1 on the FeAs layers, the FeAs equilibrium distance is 8.3 pm bigger than FeAs equilibrium distance for systems with charge +0.5 on the FeAs layers. Above a critical ionic radius the distance between the curves is 8.3 pm and this value is constantly increasing the ionic radius.

**Figure 2:**

The Fe-Fe distance is plotted as a function of the FeAs layer charge for 122 family in FeAs compounds [24, 25]. The bold line indicates the variation of the FeAs bond equilibrium distance as a function of the charge in the layer.

**Figure 3:** The microstrain is plotted as a function of the FeAs layer charge for 122 family in FeAs compounds [24, 25]. The shaded area indicates the maximum $T_c$ region.

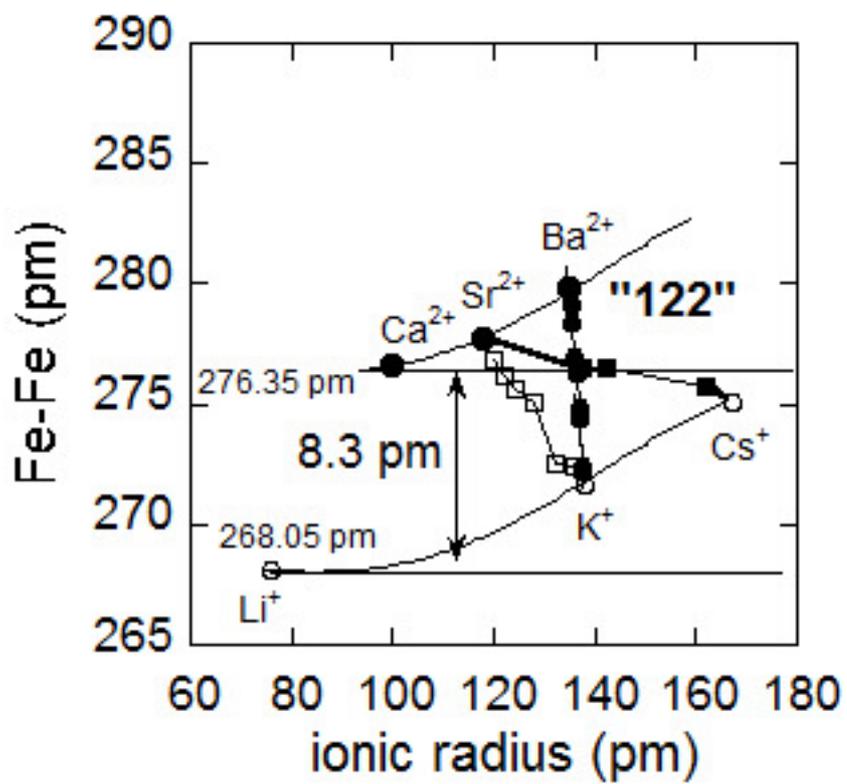

**Figure 1**

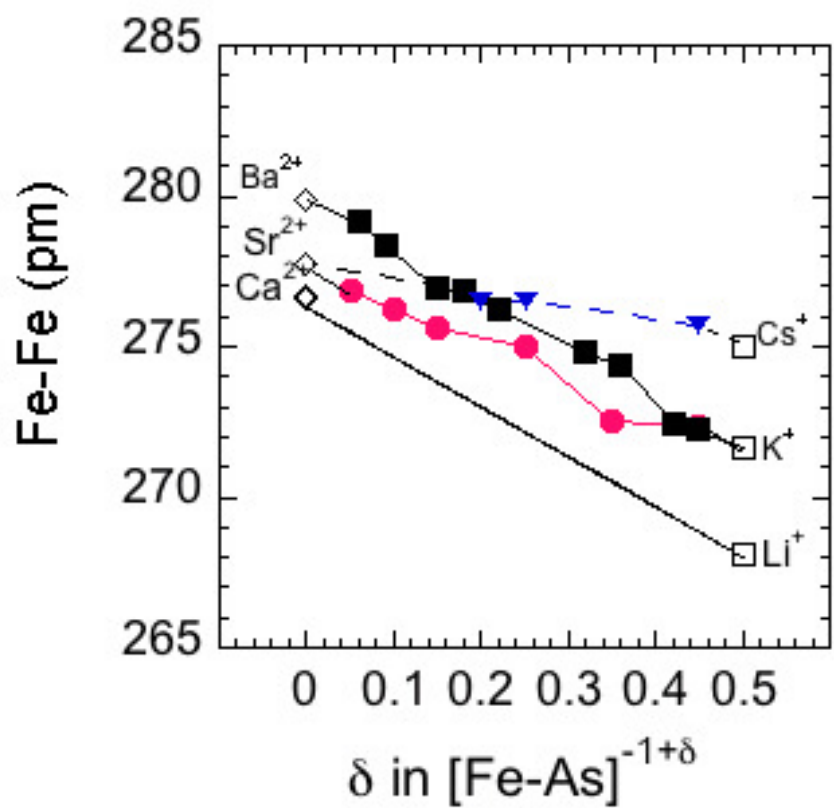

**Figure 2**

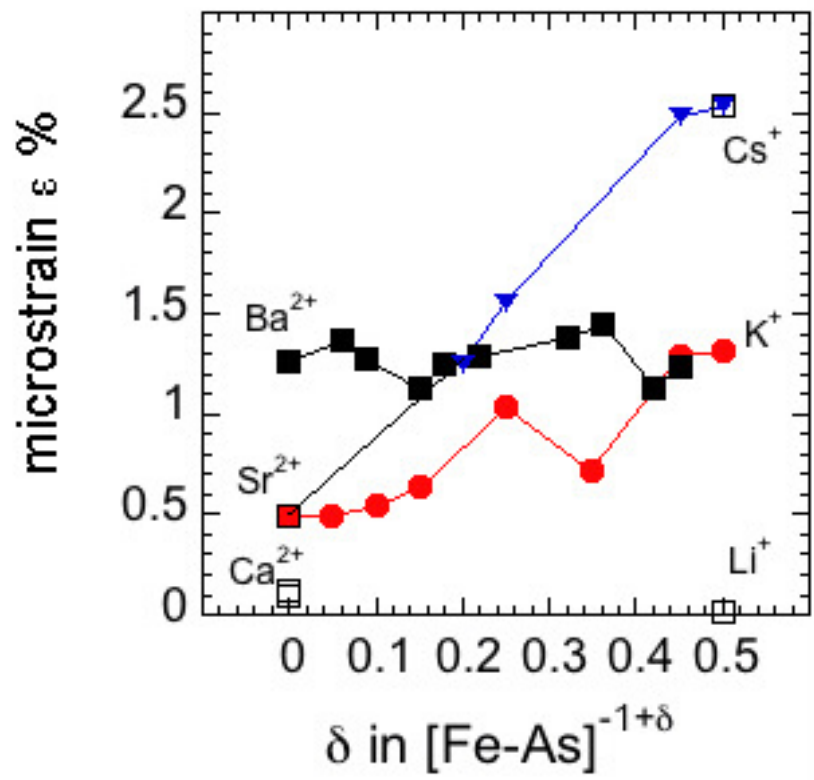

**Figure 3**